\documentclass[showpacs,twocolumn,prhrenumbers,amsmath,amssymb]{revtex4}
\usepackage{amsmath,amsfonts,latexsym,amssymb,graphicx,graphics,epsfig,subfigure,color,makeidx}
\usepackage{xcolor,epstopdf}
\usepackage{multirow}
\usepackage[colorlinks=true,citecolor=blue,linkcolor=red,anchorcolor=green,urlcolor=cyan]{hyperref}
\newcommand {\nn}{\nonumber}
\makeatletter
\providecommand \doibase [0]{http://dx.doi.org/}

\begin{document}
\title{The shadows of accelerating Kerr-Newman black hole and constraints from M87*}

\author{Tao-Tao Sui$^{a}$\footnote{suitt14@lzu.edu.cn, corresponding author}}
\author{Qi-Ming Fu$^{b,c}$\footnote{fuqiming@snut.edu.cn}}
\author{Wen-Di Guo$^{d}$\footnote{guowd@lzu.edu.cn}}

\affiliation{$^{a}$College of Physics, Nanjing University of Aeronautics and Astronautics, Nanjing 211106, China\\
$^{b}$Institute of Physics, Shaanxi University of Technology, Hanzhong 723000, China\\
$^{c}$Department of Physics, College of Sciences, Northeastern University, Shenyang 110819, China\\
$^{d}$Institute of Theoretical Physics $\&$ Research Center of Gravitation, Lanzhou University, Lanzhou 730000, China}

\begin{abstract}
In this paper, we study the influence of the parameters for the accelerating Kerr-Newman black hole on the shadows and the constraints, extensively. We find that the rotating parameter $a$, the charge parameter $e$, and the inclination angle $\theta_0$ affect the shadow qualitatively similar to that of Kerr-Newman black holes. The result shows that the size of the shadow will scale down with the accelerating factor $A$. Besides, the factor $A$ also can affect the best viewing angles, which make the observations maximum deviate from $\theta_0=\frac{\pi}{2}$, and the degree of the deviations are less than $1\%$. Then, we assume the M87* as an accelerating Kerr-Newman black hole with the mass $M=6.5\times10^9M_\odot$ and the distance $r_0=16.8Mpc$. Combining the EHT observations, we find that neither the observations, circularity deviation $\Delta C$ or axial ratio $D_x$ can distinguish the accelerating black hole or not. However, the characteristic areal-radius of the shadow curve $R_a$ can give corresponding constraints on the parameters of the accelerating Kerr-Newman black hole. The results shows that the bigger accelerating factor $A$ is, the stronger constraints on the rotating parameter $a$ and charged parameter $e$. {The maximum range of the accelerating factor is $Ar_0\leq0.558$ for a accelerating Schwarzschild case with $(a/M=e/M=0)$, and for an extremely slow accelerating case $(Ar_0\leq0.01)$, the ranges of rotating parameter $a$ and charged parameter $e$ are $a/M\in(0,1)$ and $e/M\in(0,0.9)$}.

\end{abstract}
\maketitle

\section{Introduction}\label{sec1}
When the photons move around a black hole, they will face three kinds of destinies, i.e., absorbed by the black hole, reflected by the black hole, or revolving around the black hole. {For the third destiny, the photons are trapped in unstable circular geodesics, defining the photon sphere. The photon sphere can describe the outline of the black hole, also named with the shadow of the black hole. Besides, perturbations within the photon sphere give rise to the familiar excitations known as photon sphere quasinormal modes \cite{Cardoso:2008bp,Cardoso:2017soq}.}

In the 1960s, Synge calculated the shadow of a spherically symmetric Schwarzschild black hole \cite{synge1966escape}. Then, Luminet considered the case that the Schwarzschild black hole surrounded by an accretion disk, and calculated the size of the shadow \cite{luminet1979image}. In Ref. \cite{hawking1973black}, the authors studied the shadow of a rotating black hole, and the result shows that with the dragging effect of rotation, the shadow will deviate from the perfect circle. Then, the study on the shadow of different spacetime geometries has been blooming with the motivation that the non-circular shadow, including black hole \cite{Hioki:2008zw,Grenzebach:2014fha,Wang:2017qhh,Guo:2018kis,Li:2020drn,Yan:2019etp,Hennigar:2018hza,Konoplya:2019sns,Bambi:2008jg,Bambi:2010hf,Konoplya:2019fpy,Wei:2019pjf,Wei:2018xks,Wang:2018prk,Liu:2020ola,Kumar:2019pjp,Cunha:2018acu}, and wormhole \cite{Ohgami:2015nra,Nedkova:2013msa,Shaikh:2018kfv,Amir:2018szm,Amir:2018pcu,Wang:2020emr,Chang:2020lmg,Gralla:2020yvo}.

More recently, the Event Horizon Telescope (EHT) Collaboration gave the image of the supermassive black hole M87*, which is the first image of the shadow for the black hole and makes the theoretical prophecy of the black hole shadow become physical reality \cite{Akiyama:2019cqa,Akiyama:2019brx,Akiyama:2019sww,Akiyama:2019bqs,Akiyama:2019fyp,Akiyama:2019eap}. The shadow of M87* gives the observations with the derivation circularity $\Delta C\lesssim0.1$, the axis ratio $1<D_x\lesssim4/3$, and the angular shadow diameter $\psi_d=3\sqrt{3}(1\pm0.17)\psi_g$, with the angular gravitional radius $\psi_g=3.8\pm0.4\mu as$. {Combining with these observations of M87*, the authors have effectively constrained the characteristics of Kerr-like black holes \cite{Meng,Kuang:2022ojj,Xavier:2023exm}. Furthermore, various methods exist for constraining black hole properties. For instance, Kuang and Ali employ gravitational lensing to constrain Kerr-like black holes \cite{Kuang:2022xjp}, while Destounis et al. establish limitations for non-Kerr black holes \cite{Destounis:2020kss, Destounis:2021, Destounis:2021pd, Destounis:2023} as well as exotic compact objects \cite{Maggio:2021uge, Destounis:2023khj} through gravitational-wave observables. Additionally, there are endeavors to constrain non-vacuum black holes situated within astrophysical contexts \cite{Cardoso:2021wlq, Cardoso:2022whc, Destounis:2022obl, Figueiredo:2023gas}.}

According to previous works, we can find that many rotating black boles share a similar characteristic, that the inclination angle of the observer equals $\pi/2$, which makes the observations of the shadows for these black holes maximum. Besides, all these researches show that the apparent characteristics of the black hole shadows vary with the parameters of the black holes, which are closely connected with the essential properties of gravity theory and the background fields. In other words, the different characteristics of black hole shadows can reflect the different forms of black hole solutions. 

On the other hand, there is a special family of black hole solutions in the Einstein's general relativity named as C-metric \cite{Berlin,Newman,Robinson,Griffiths,Kinnersley:1970zw}, which can describe the accelerating black hole and belongs to the well-known Plebański-Demiański spacetime \cite{Plebanski:1976gy,Griffiths:2005qp}. {Many works not only focus on the thermodynamics of the C-metric \cite{Appels:2016uha,Appels:2017xoe,Astorino:2016ybm,Gregory:2017ogk}, but also on its stability, superradiance effect and the validity of strong cosmic censorship in these spacetimes \cite{Destounis1,Destounis2,Destounis3,Fontana}}. In Refs. \cite{Grenzebach:2015oea,Zhang:2020xub}, the authors just analyzed the shadows of the accelerating Kerr black holes theoretically. The result shows that the acceleration of the black hole has an unique influence on the properties of the shadow.

In Ref. \cite{Anabalon:2018qfv}, the authors obtained the consistent thermodynamic description of the accelerating charged and rotating black hole, which also can be named as accelerating Kerr-Newman (KN) black hole. Inspired by the previous works \cite{Grenzebach:2015oea,Zhang:2020xub}, we can not only investigate the influence of the parameters, i.e. the rotating parameter, the charged parameter, and the accelerating parameter on the shadow for the accelerating KN black hole, and analyze the characterized observables of the black bole shadow, theoretically. But also, we can consider the supermassive black hole M87* as the accelerating KN black hole and constrain the corresponding parameters with the EHT observations.

The main parts of this paper are organized as follows. In Sec. \ref{sec2}, we will investigate the circular orbits of the photons around the accelerating Kerr-Newman black hole. In Sec. \ref{sec3}, we will consider the influence of the parameters on the shadow for the accelerating Kerr-Newman black hole. We presuppose the M87* as the accelerating Kerr-Newman black hole, and get the constraints of the corresponding parameters from the EHT observations in Sec. \ref{sec5}. The last section contributes to our closing summary. 

\section{Circular photon orbits around the accelerating Kerr-Newman black hole}\label{sec2}
In this section, we first give a brief review of the accelerating KN black hole, the line element can be described as \cite{Anabalon:2018qfv}
\begin{align}\label{metric}
d s^{2}=&\frac{1}{H^2}\Big\{-\frac{\Delta_r}{\Sigma}\big(\frac{dt}{\alpha}-a \sin^2\theta d\phi\big)^2+\frac{\Sigma}{\Delta_r}dr^2+\frac{\Sigma}{\Delta_\theta}d\theta^2\nonumber\\
&+\frac{\Delta_\theta}{\Sigma}\Big(\frac{a dt}{\alpha}-(r^2+a^2)d\phi\Big)^2\Big\},
\end{align}
with
\begin{equation}\begin{aligned}
H&=1+Ar\cos\theta,~\Sigma=r^2+a^2 \cos^2\theta,\\ 
\Delta_{r}&=(1-A^2r^2)(r^{2}-2 M r+a^{2}+e^2),\\
\Delta_{\theta}&=1+2MA\cos\theta+A^2(a^2+e^2)\cos^2\theta,\\
\alpha&=\frac{\sqrt{(1-a^2A^2)(1+a^2A^2+e^2A^2)}}{1+a^{2} A^{2}},
\end{aligned}\end{equation}
and the corresponding gauge potential can be expressed as
\begin{equation}\begin{aligned}
&F=dB, \\
&B=-\frac{e}{\Sigma r}\big(\frac{dt}{\alpha}-a\sin^2\theta d\phi\big)+\frac{e r_{+}}{(a^2+r_{+}^2)\alpha}dt,
\end{aligned}\end{equation}
{where $r_{+}$ is the radius of the event horizon of the accelerating KN black hole with $\Delta_r(r_+)=0$.}

In the charged accelerating Kerr black hole, the parameters $m$, $a$ and $e$ are the mass, rotation and charge parameters, respectively. The parameter $A$ represents the acceleration factor of the black hole. $H$ is the conformal factor, which can affect the conformal boundary $r_{B}$ of the accelerating black hole with $r_{B}=\frac{1}{|A\cos \theta|}$. The parameter $\alpha$ can rescale the time coordinate and ensure the Killing vector be normalized at conformal infinity. {The term $\Delta_r$ can also be expressed as $\Delta_r=(r-r_{-})(r-r_{+})(r^2-r^2_{A})$, where $r_{\pm}$ are identical to the event and Cauchy horizons of the non-accelerating KN black hole, $r_{A}=1/A$ is already familiar in the context of the C-metric as an acceleration horizon. With the conformal boundary $r_{B}$, the range of $r$ should be constrained as $r_{+}\leq r\leq r_{B}$. Besides, in order to make sure that the accelerating KN black hole in the bulk, the term $\Delta_r=0$ should have roots in the range $r<r_{A}$ \cite{Anabalon:2018qfv}.}  

Then, the Lagrange for the photons can be described by
\begin{equation}
\mathcal{L}=\frac{1}{2}g^{\mu\nu}\dot{x}_\mu\dot{x}_\nu,
\end{equation}
with the definition $\dot{x}^{\mu}=dx^{\mu}/d\lambda=u^{\mu}$, where $u^{\mu}$ is the four-velocity of the photon and the parameter $\lambda$ is the affine parameter. {For this stationary black hole, there are two Killing vectors $\partial_{t}$ and $\partial_{\phi}$ which can result the conserved total energy $E$ and z-component of the angular momentum $L_z$.}

Using the Hamilton-Jacobi equation, we can get the null geodesic equation of the photon on the background of the accelerating KN black hole 
\begin{align}
\frac{\Sigma}{H^{2}}\frac{dt}{d\lambda}=& \frac{\alpha\left(a^2+r^2\right) \left(\alpha E\left(a^2+r^2\right)-a L_z\right)}{\Delta _r} \nonumber \label{teff}\\
&+\frac{a \alpha\left(L_z-a \alpha E\sin ^2\theta \right)}{\Delta _{\theta }},\\
\frac{\Sigma}{H^{2}}\frac{d\phi}{d\lambda}=& \frac{a \left(\alpha E\left(a^2+r^2\right)-a L_z\right)}{\Delta _r}\nonumber\\
&+\frac{\left(L_z \csc ^2 \theta -a \alpha E\right)}{\Delta _{\theta }}, \label{phieff}\\
\left(\frac{\Sigma}{H^{2}}\right)^{2}\left( \frac{dr}{d\lambda}\right)^2 =&\big[(a^2+r^2) E-a L_z\big]^2-\Delta _r Q\equiv R(r),\label{reff}\\
\left(\frac{\Sigma}{H^{2}}\right)^{2}\left( \frac{d\theta}{d\lambda}\right)^2=&\Delta _{\theta}Q-\frac{(L_z-a E \sin^2\theta)^2}{\sin^2\theta}\equiv\Theta(\theta),\label{loeff}
\end{align}
where parameter $Q$ is the Carter constant \cite{Carter:1968rr} and the functions $R(r)$ and $\Theta(\theta)$ can be considered as the radial and longitudinal effective potential, respectively.

Since the photon orbits should independent of the energy, we can introduce the following dimensionless abbreviations to describe the photon orbits 
\begin{equation}
\xi=\frac{L_z}{E},\quad \eta=\frac{Q}{E^2}.
\end{equation}
Generally, the unstable spherical circular photon orbits should satisfy with the conditions
\begin{equation}
R(r)=0,\quad \frac{d R(r)}{dr}=0.\label{co1}
\end{equation}
According to the above two conditions, the dimensionless quantities $\xi$, and $\eta$ can be solved as
\begin{align}
\xi&=\frac{a^2 \partial_r\Delta_r+r^2\partial_r\Delta_r-4 r \Delta_r}{a~ \partial_r\Delta_r},\label{xi}\\
\eta&=\frac{16 r^2 \Delta_r}{(\partial_r\Delta_r)^2}.\label{eta}
\end{align}

By plugging these expressions into  Eq.\eqref{loeff}, we can find that the non-negativity of Eq. \eqref{loeff} gives the condition for the photon region as 
\begin{equation}
(4r\Delta_r-\Sigma\partial_r\Delta_r)^2 \leq 16a^2r^2\Delta_r\Delta_\theta\sin^2\theta,
\end{equation}
For the non-rotating case $a=0$, the circular orbit radius can be solved from the unstable spherical circular photon orbits conditions \eqref{co1} as
\begin{equation}
r_p =\frac{-1+A^2 e^2}{3 A^2 M}+2 \sqrt{\beta} \cos \Big[\frac{1}{3}\cos ^{-1}(\alpha\beta^{-\frac{3}{2}})\Big],\label{sc1}
\end{equation}
with 
\begin{align}
\alpha&=\frac{2 \left(A^2 e^2-1\right)^3-27 A^2 M^2 \left(A^2 e^2+1\right)}{54 A^6 M^3},\nonumber\\
\beta&=\frac{1}{A^2}+\frac{\left(A^2 e^2-1\right)^2}{9 A^4 M^2}.
\end{align}

\section{The shadows of the charged accelerating Kerr black holes}\label{sec3}
 
For calculating the character of the black hole shadow, which can be seen by an observer, we should assume a normalized and orthogonal frame where the observer is located at, and this frame can be expressed as following
\begin{align}
\hat{e}_{(t)}=&\sqrt{\frac{g_{\phi \phi}}{g_{t \phi}^2 - g_{t t} g_{\phi\phi}}} \left( \partial_t - \frac{g_{t \phi}}{g_{\phi \phi}} \partial_{\phi}\right),\label{et}\\
\hat{e}_{(r)}=& \frac{1}{\sqrt{g_{r r}}} \partial_r,~\hat{e}_{(\theta)}=\frac{1}{\sqrt{g_{\theta \theta}}} \partial_{\theta},~\hat{e}_{(\phi)}=\frac{1}{\sqrt{g_{\phi \phi}}} \partial_{\phi}.
\end{align}
In this frame, we can see that the observer is locally static and zero angular momentum with respect to infinity for $\hat{e}_{(t)}\cdot\partial_{\phi}=0$. Hence, this frame also can be named as zero-angular-momentum-observer (ZAMO) reference frame.  

For describing the shadow of the accelerating KN black hole, we should project the 4-momentum $p^{\mu}$ of photon onto the ZAMO reference frame where the observer located at, and the corresponding quantities measured by the observer can be expressed as
\begin{equation}
p^{(t)} =- p_{\mu} \hat{e}_{(t)}^{\mu}, 
\quad p^{(i)}= p_{\mu} \hat{e}_{(i)}^{\mu}, ~~i = r, \theta, \phi.
\end{equation}
In the ZAMO frame, the spatial component of momentum $|\vec{P}|$ should satisfy $|\vec{P} | = p^{(t)}$ for the massless photon. In order to express the specific 3-momentum $p^{i}$, we can introduce the observation angles $(\alpha,\beta)$ as \cite{Cunha:2016bpi}
\begin{eqnarray}
p^{(r)} & = & | \vec{P} | \cos \alpha \cos \beta, ~p^{(\theta)} = | \vec{P} | \sin \alpha,  \nn \\
p^{(\phi)} & = & | \vec{P} | \cos \alpha \sin \beta.
\end{eqnarray}
Combining the geodesic equations \eqref{teff}-\eqref{loeff} together, we can obtain
\begin{eqnarray}
\sin\alpha&=&\frac{p^{(\theta)}}{p^{(t)}}=\pm\frac{H}{\zeta-\gamma\xi}\nonumber\\
&&\times\sqrt{\frac{\eta}{\Sigma}-\frac{(a\sin^2\theta-\xi)^2}{\Sigma \Delta_\theta\sin^2\theta}}|_{(r_O,\theta_O)},\\
\tan\beta&=&\frac{p^{(\phi)}}{p^{(r)}}=\frac{\xi\sqrt{\Sigma\Delta_r}}{H\sqrt{g_{\phi\phi}}}\nonumber\\
&&\times\Big[(r^2+a^2-a\xi)^2-\Delta_r\eta\Big]^{-\frac{1}{2}}|_{(r_O,\theta_O)}.
\end{eqnarray}
Here, to simplify the above expression, we set the abbreviations  $\zeta\equiv\hat{e}_{(t)}^t$, and $\gamma\equiv\hat{e}_{(t)}^\phi$. The parameters $r_{O}$ and $\theta_{O}$ are the radial position and the inclination angle between the observer and the direction of the rotation axis for the black hole.

\subsection{The Shadows}

\begin{figure}[htbp!]
\begin{center}
\subfigure[$\ a$]{\label{aeffect}
\includegraphics[width=0.232\textwidth]{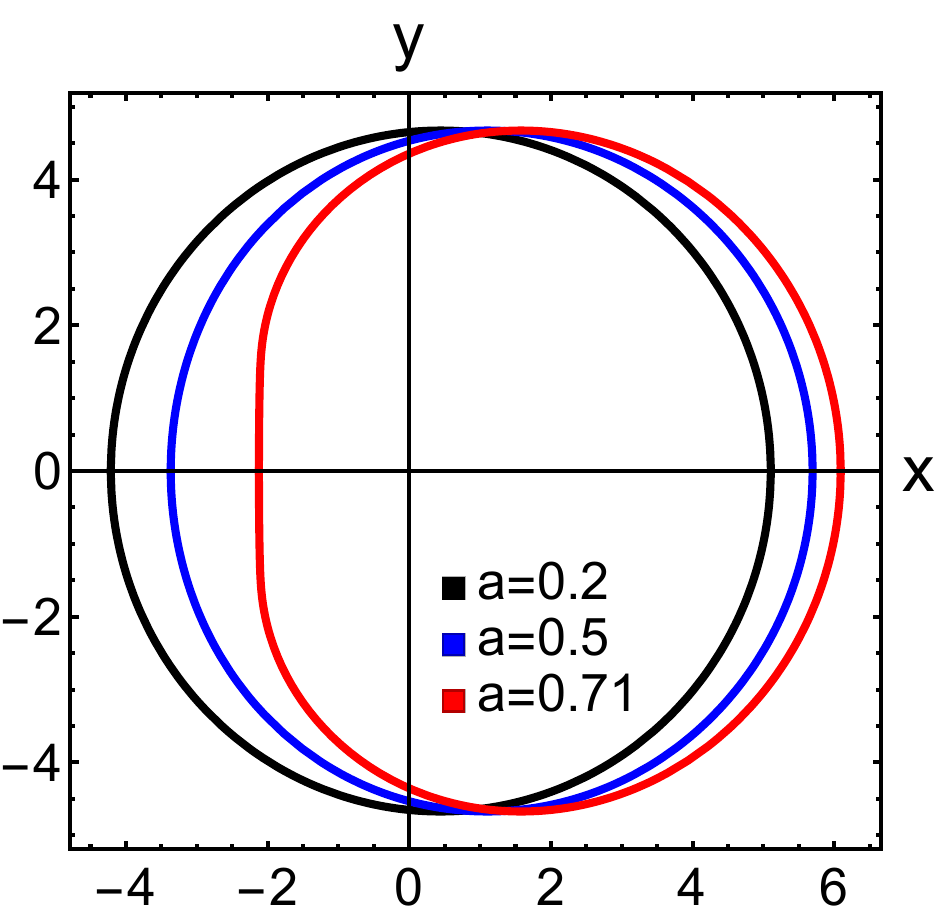}}
\subfigure[$\ \theta_{O}$]{\label{psieffect}
\includegraphics[width=0.232\textwidth]{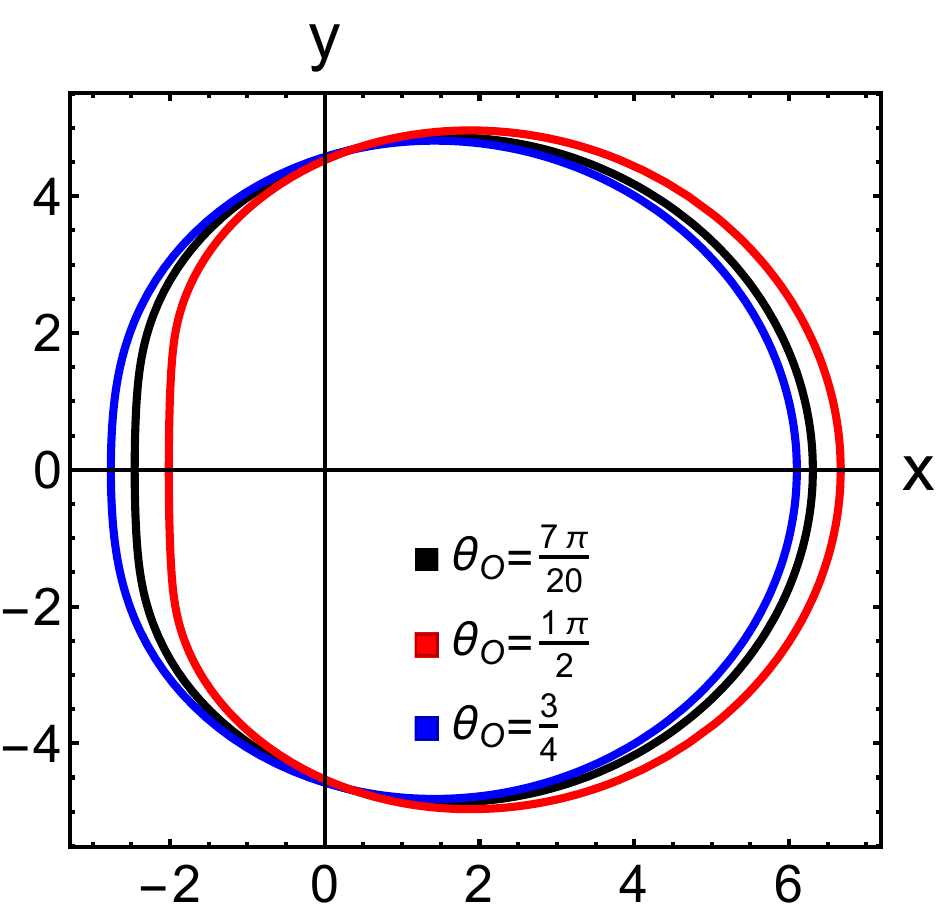}}
\end{center}
 \caption {The examples of shadows for the accelerating KN black holes with different parameters. The other parameters are set as $\theta_{O}=\frac{\pi}{2}, e=0.4, A=0.002$ for (a) and $e=0.4, a=0.9, A=0.002$ for (b).}\label{f1}
\end{figure}

\begin{figure}[htbp!]
\begin{center}
\subfigure[$~e$]{\label{eeffect}
\includegraphics[width=0.232\textwidth]{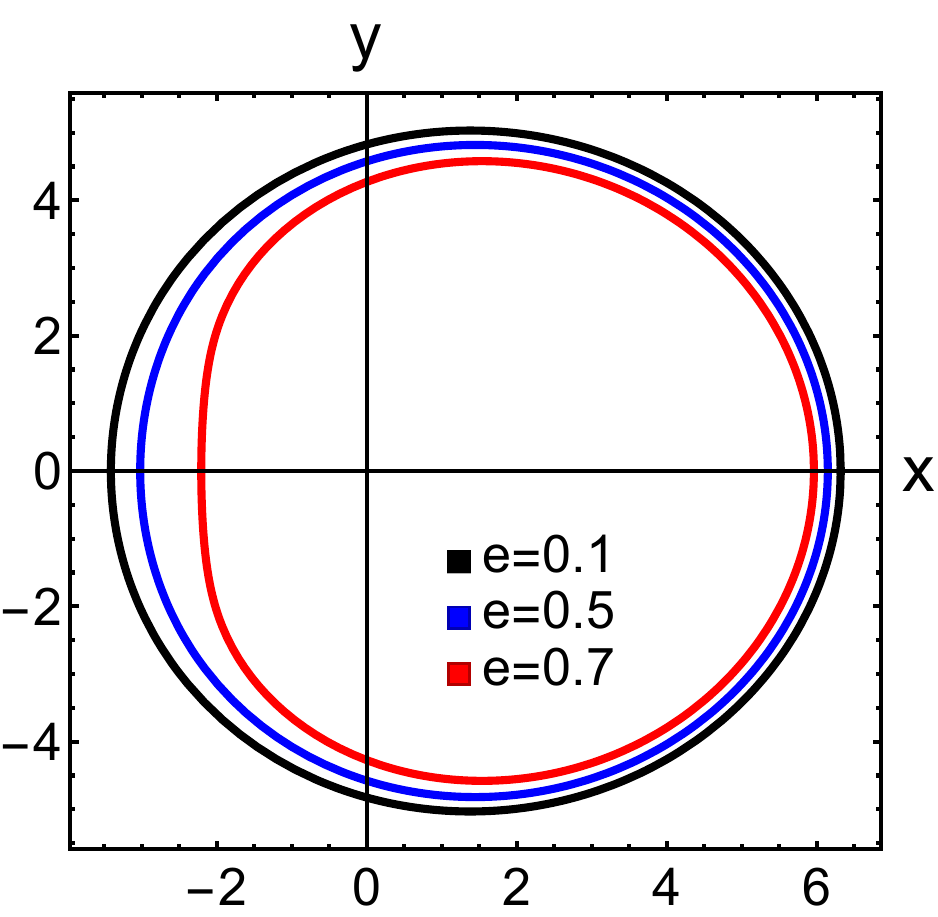}}
\subfigure[$~A$]{\label{thffect}
\includegraphics[width=0.23\textwidth]{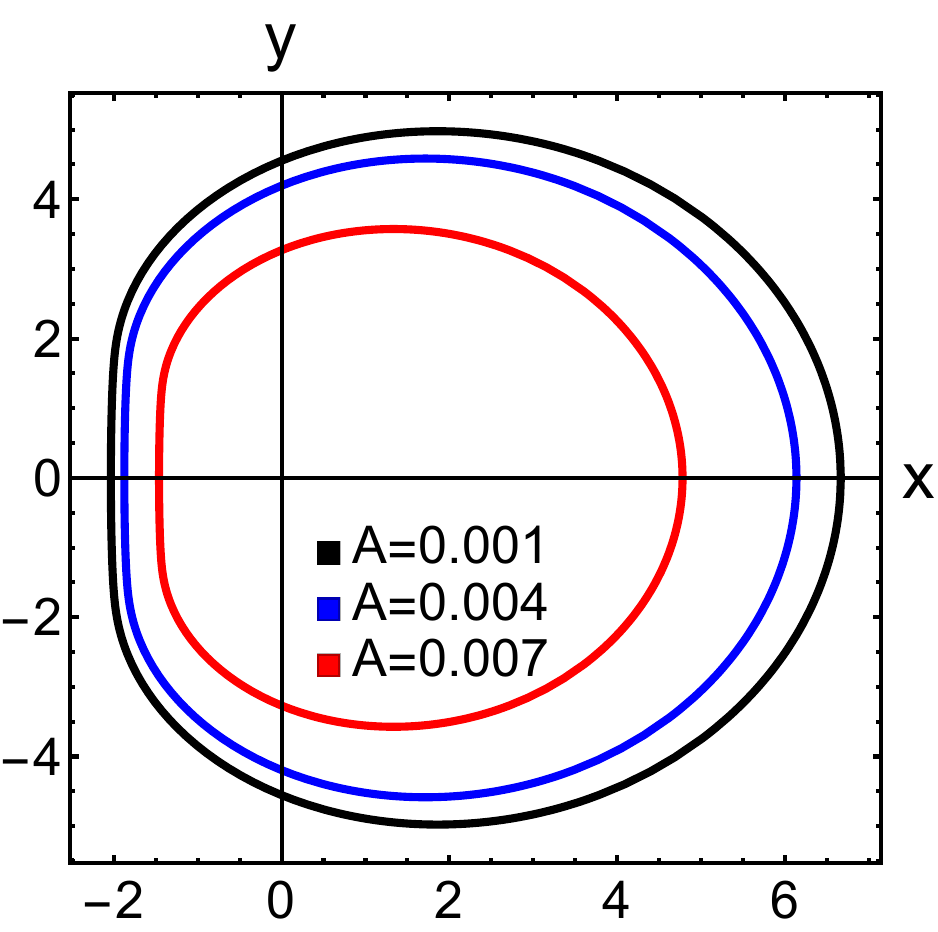}}
\end{center}
\caption {The patterns of shadows of the accelerating KN black holes with different parameters. The other parameters are set as $\theta_{O}=\frac{\pi}{2}, a=0.7, A=0.002$ for (a) and $\theta_{O}=\frac{\pi}{2}, a=0.7, e=0.7$ for (b).} \label{f2}
\end{figure}

Furthermore, in order to obtain the apparent position on the plane of the sky for the observer, we need to introduce the Cartesian coordinate $(x, y)$ as
\begin{eqnarray}
x \equiv - r_O \beta, \quad y \equiv r_O \alpha.\label{xy}
\end{eqnarray}
We can see that both the coordinates $x$ and $y$ are the functions of $(r_0, \theta_O, a, e, A)$. We show the shadows of the accelerating KN black hole for the variables $a$, $e$, $A$ and $\theta_O$ in Figs. \ref{f1} and \ref{f2}, where we set $M=1$ and $r_O=100$. From Fig. \ref{aeffect}, we can see that with the increasing of the angular momentum $a$, the shape of the rotating black hole shadow becomes non-circular as the result due to the dragging effect. Besides, Fig. \ref{psieffect} shows that the observable dragging effect which can result in the shadows of rotating black hole deviating from the standard circle also depends on the observation angle $\theta_{O}$, even though the charged accelerating Kerr black hole with large rotating parameter. From Fig \ref{f2}, we should note that the size of the black hole shadow decreases with the charge parameter $e$ and the acceleration factor $A$. At the same time, as the charge parameter $e$ increases, the non-circular behavior of black hole shadows becomes more significant. Summing up, we can draw a conclusion  that the phenomena for the shadow of Kerr black hole deviating from the standard circle are mainly caused by the rotating of the black hole, and the presence of charge of the black hole will assist in the generation of the phenomena, which is similar to that of the Kerr-Newman cases \cite{Perlick,Grenzebach,Tsukamoto,Xavier}. Furthermore, the acceleration factor $A$ can have an effect on the size of the shadow for the black hole, and we will further investigate other influences of the factor $A$ on the shadow in the next part.

\subsection{The Observables}\label{sec4}

In the previous section, we analyzed the influence of the parameters for the accelerating KN black holes on their shadows. However, we only can get the shadows of the black hole by astronomical observation. So, in order to know the characters of the accelerating KN hole by the astronomical observed data, we should construct some easily measured and reliable astronomical observables. Here, we will consider the following observables with the radius $R_s$ and the distortion parameter $\delta_{s}$, which are defined by Hioki and Maeda \cite{Hioki:2009na}. The parameter $R_{s}$ can be considered as a radius of the reference circle which should pass the top point $(x_t,y_t)$, the bottom point $(x_b,y_b)$, and the right point $(x_r,0)$ of the shadow, and the distortion parameter $\delta_{s}$ can measure the distortion of the black hole shadow compared with the reference circle, which can be defined as
\begin{eqnarray}
R_{s}=\frac{(x_{t}-x_{r})^{2}+y_{t}^{2}}{2(x_{r}-x_{t})},~~
\delta_{s}=\frac{x_{l}-\tilde{x}_{l}}{R_{s}}.
\end{eqnarray}
where $(x_{l},0)$ and $(\tilde{x}_{l},0)$ are the left points of the shadow and reference circle, respectively. 

\begin{figure}[htbp!]
\begin{center}
\subfigure[]{\label{Rae}
\includegraphics[width=0.232\textwidth]{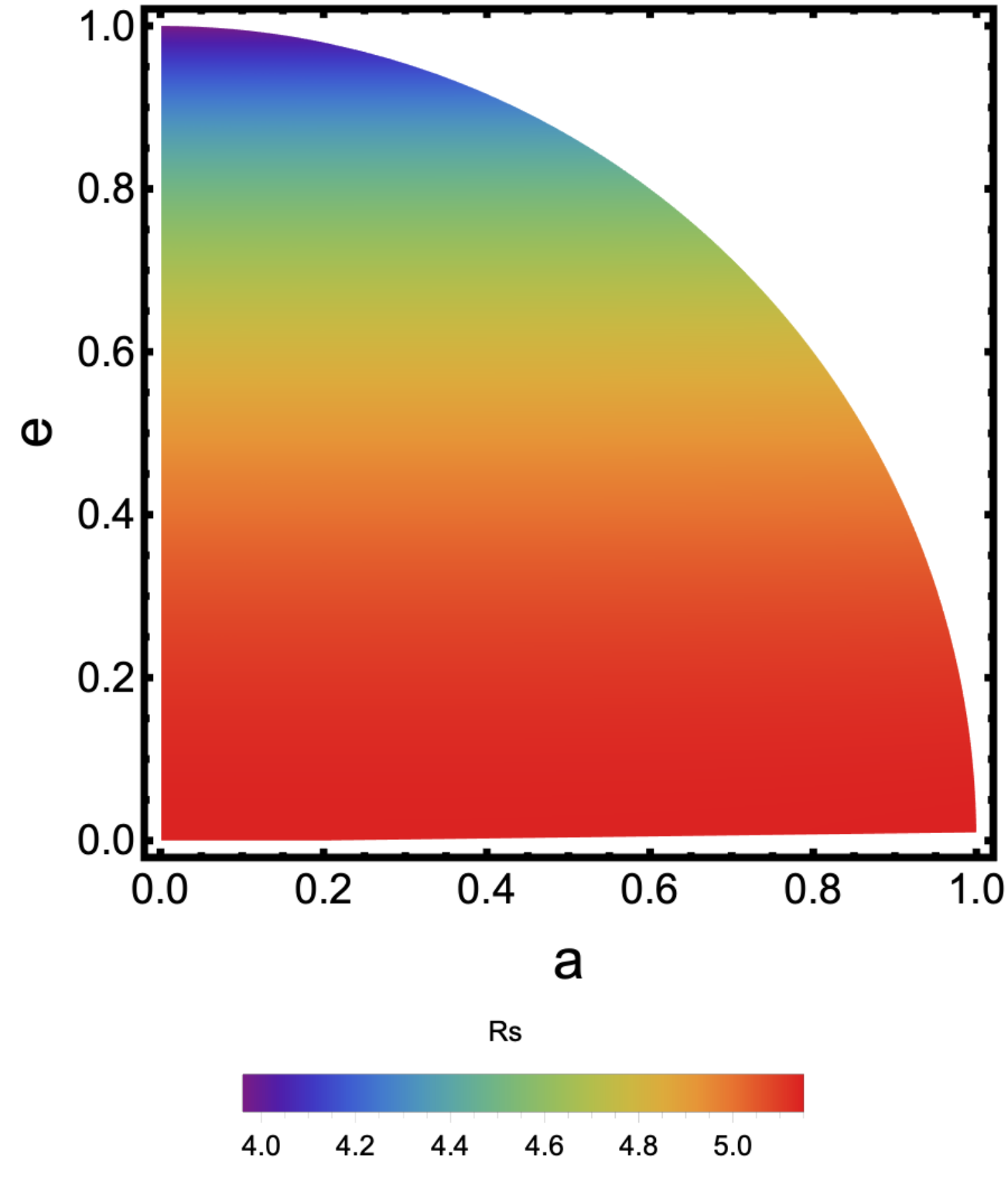}}
\subfigure[]{\label{dae}
\includegraphics[width=0.232\textwidth]{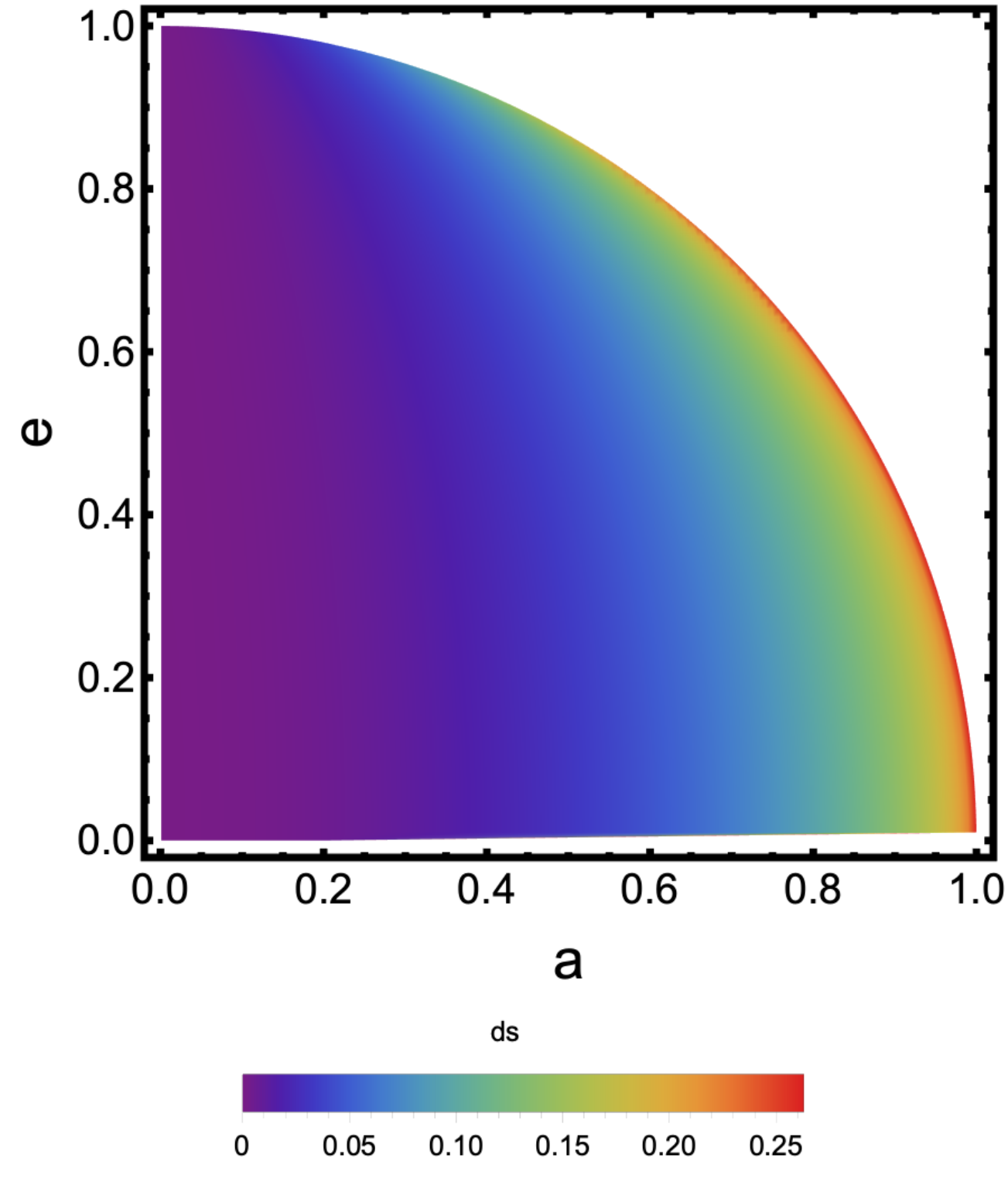}}
\end{center}
\caption {The density plots of the observable parameters $R_{s}$ and $\delta s$ with the variables ($a$,$e$). The other parameters are set as $\theta_{O}=\frac{\pi}{2}$, $r_0=100$, and $A=0.005$.} \label{f3}
\end{figure}

Figure \ref{f3} displays the influence of the rotation parameter $a$ and charge parameter $e$ on the observable parameters $R_{s}$ and $\delta_s$. From Fig. \ref{Rae}, we can see that the radius parameter $R_{s}$ decreases rapidly with the increasing charge parameter $e$. While, the rotating parameter $a$ slightly affect the reference circle $R_{s}$. For the distortion parameter $\delta_s$, Fig. \ref{dae} shows that the distortion of the shadow for the accelerating KN black hole increases with the rotation parameter $a$, and the effect of charged parameter $e$ is slight. The effects of the rotating parameter $a$ and the charged parameter $e$ on the astronomical observables $R_{s}$ and $\delta_s$, are  similar to that of Kerr-Newman black hole \cite{Kumar:2019pjp}.

\begin{figure}[htbp!]
\begin{center}
\subfigure[]{\label{RApsi}
\includegraphics[width=0.232\textwidth]{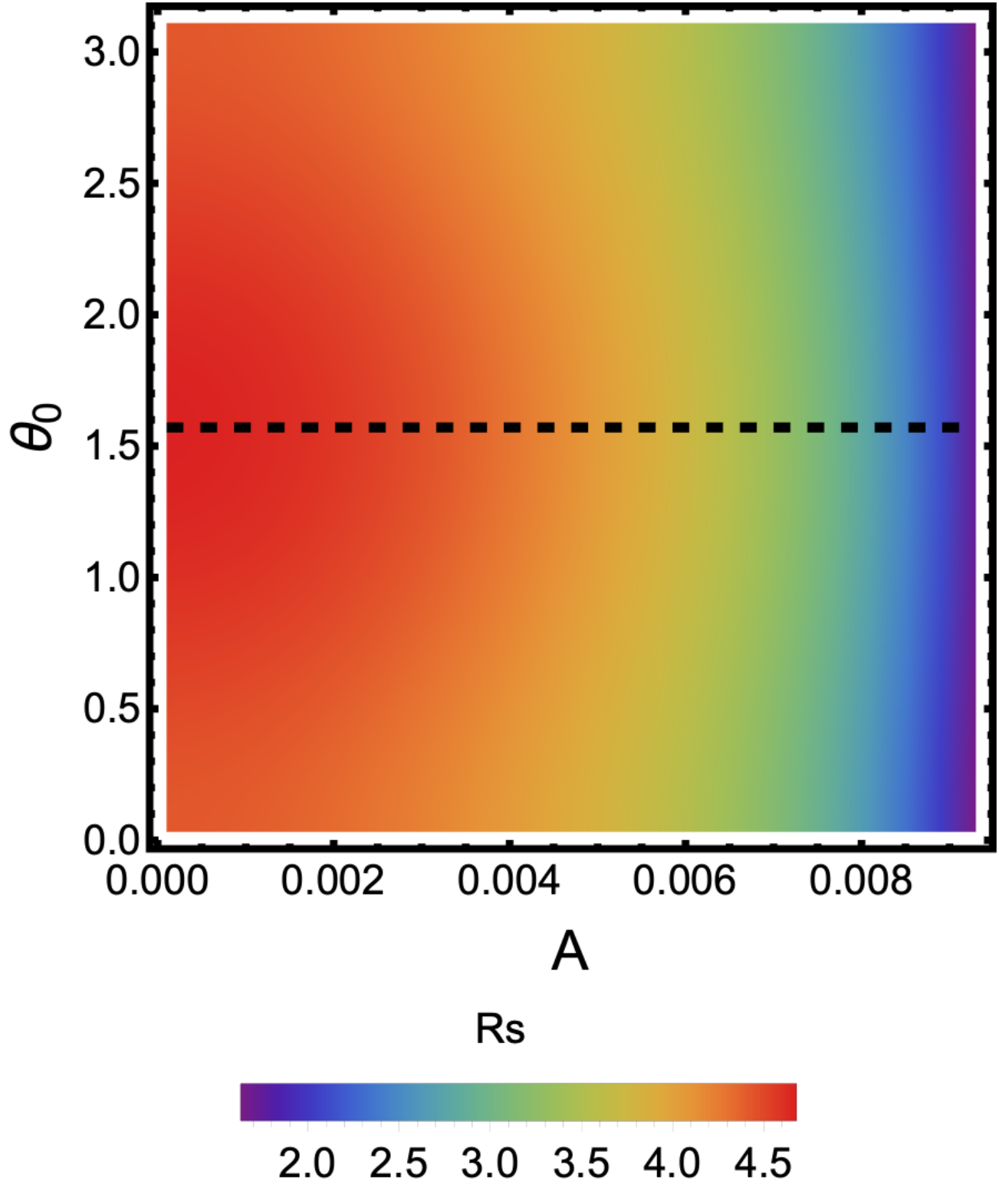}}
\subfigure[]{\label{eApsi}
\includegraphics[width=0.232\textwidth]{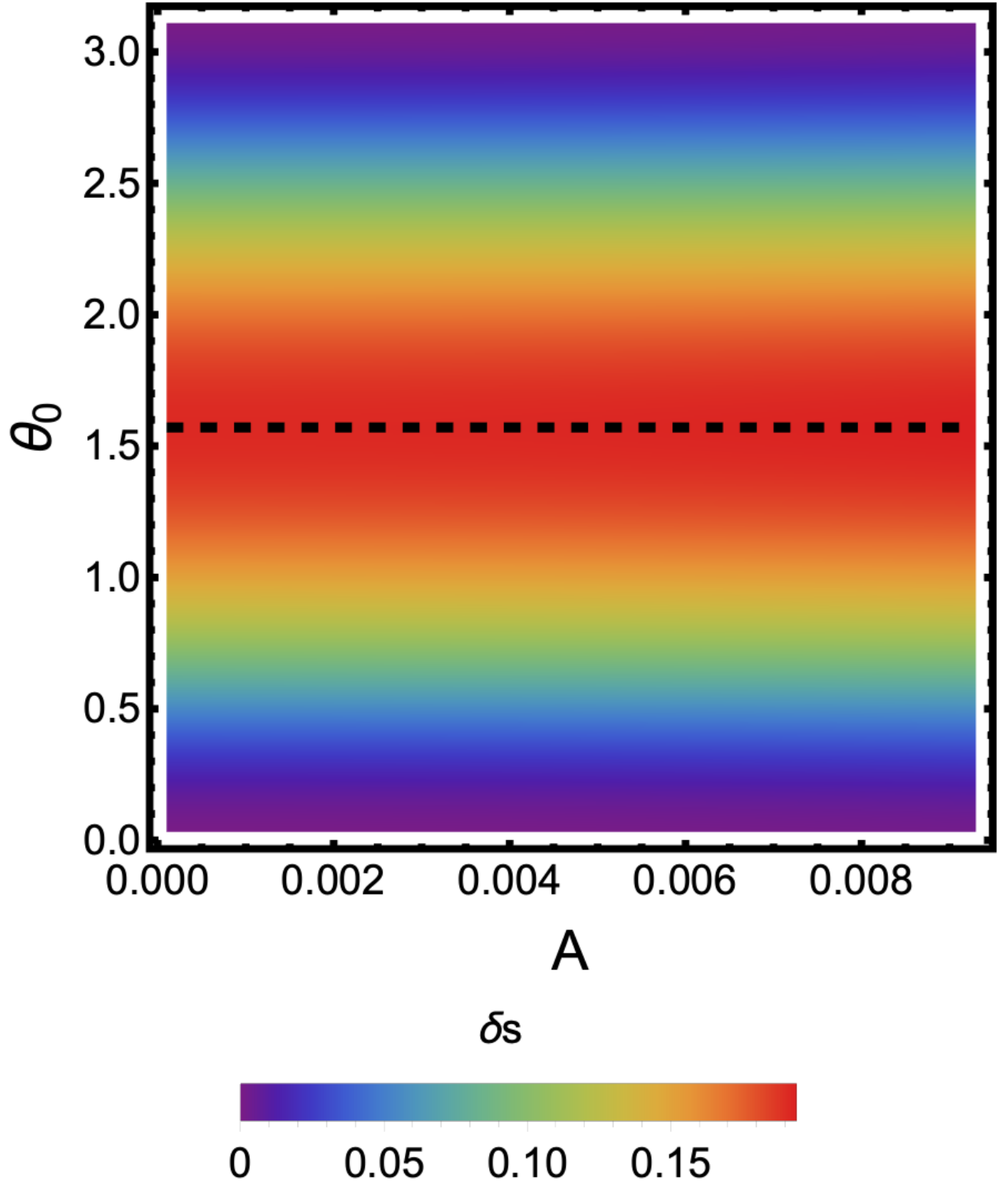}}
\end{center}
\caption {The density plots of the observable parameters $R_{s}$ and $\delta s$ with the variables ($A$,$\theta_0$). Here, the other parameters are set as $r_0=100$,  $a=0.7$, and $e=0.7$. The dashed black lines denote $R_{s}$ and $\delta s$ with $\theta_0=\frac{\pi}{2}$.} \label{f4}
\end{figure}

In previous chapter, the result shows that the acceleration factor $A$ can affect the size of the shadow of the black hole, which can also be verified in Fig. \ref{RApsi}. From Fig. \ref{f4}, we can see that the observation angle $\theta_O$ results a slighter influence on the radius $R_s$, but a larger effect on the distortion $\delta_s$. Furthermore, when the inclination angle $\theta_0\approx\frac{\pi}{2}$, both $R_s$ and $\delta_s$ reach the maximum value, with other parameters are fixed.

\begin{figure}[htbp!]
\begin{center}
\subfigure[]{\label{thetar}
\includegraphics[width=0.40\textwidth]{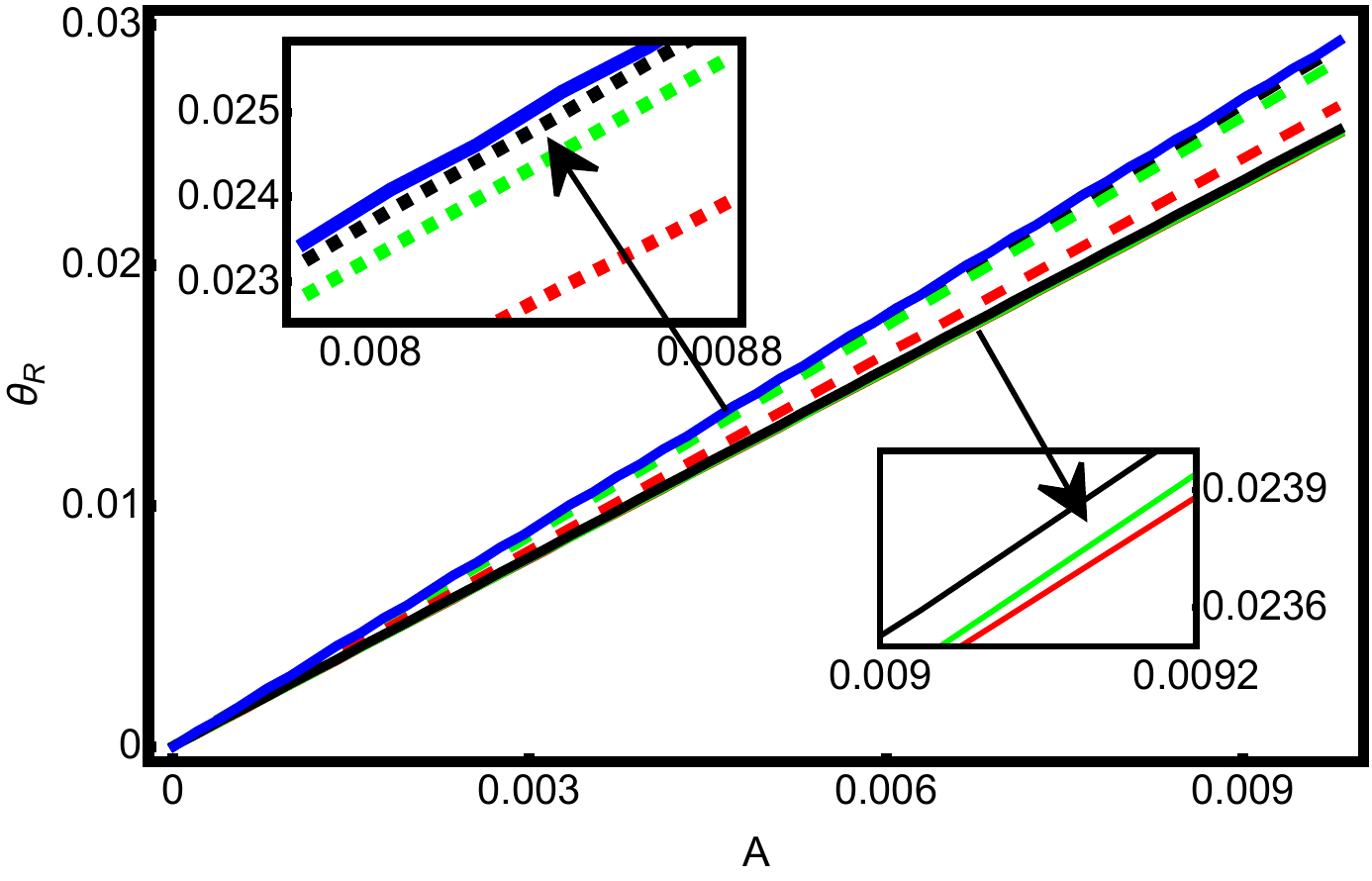}}\\
\subfigure[]{\label{thetad}
\includegraphics[width=0.40\textwidth]{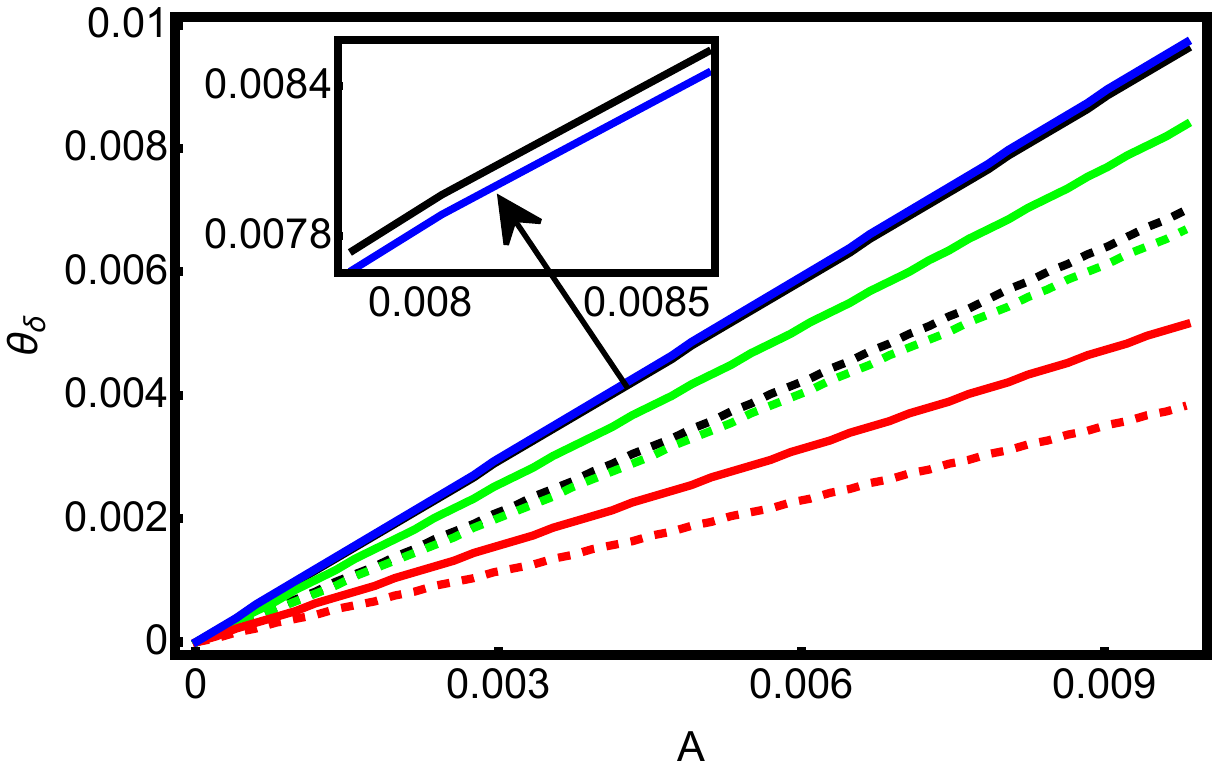}}
\end{center}
\caption {The diagrams for the deviation of the inclination angle which makes the shadows radius $R_s$ and the distortion $\delta_s$ maximum. Here, we set $M=1$ and $r_0=100$. The dashed lines are black ($e=0.2$), green ($e=0.4$), and red ($e=0.59$) with $a=0.8$. The solid lines are black ($a=0.2$), green ($a=0.4$), and red ($a=0.7$) with $e=0.7$. Besides, the solid blue line is ($a=0.05,e=0$).} \label{f5}
\end{figure}

In order to ascertain the influence of the acceleration factor $A$ on the best viewing angle $\theta_0$ which makes the radius $R_s$ or $\delta s$ maximum, we can introduce the deviation angle $\theta_R$ and $\theta_s$ with the definition as \cite{Zhang:2020xub} 
\begin{eqnarray}
\theta_{R}&=&\frac{\pi}{2}-\theta_b|_{(R_s=R_{max})},\\
\theta{s}&=&\frac{\pi}{2}-\theta_b|_{(\delta_s=\delta_{s_{max}})}.
\end{eqnarray}
Figure \ref{f5} shows the variation of the degree of deviation angles $\theta_R$ and $\theta_s$ as the result of the accelerating factor $A$. For the deviation angle $\theta_R$, Fig. \ref{thetar} shows that the deviation angle $\theta_R$ decreases with the charged parameter $e$, while the rotating parameter $a$ has no effect on $\theta_R$. From Fig. \ref{thetad}, we can see that the deviation angle $\theta_s$ decreases with both the rotating parameter $a$ and charged parameter $e$. Besides, both the deviation angles $\theta_R$ and $\theta_s$ increase with the accelerating factor $A$. While, the the accelerating factor $A$ has a greater influence on the deviation angle $\theta_R$ than on $\theta_s$, and the degree of the deviations are less than $1\%$.


\section{Constraints of the parameters from M87*}\label{sec5}
The shadow image of the supermassive black hole M87* as crescent shaped was  photographed by the EHT collaboration with the mass $M=6.5\times10^9M_\odot$, the distance $r_0=16.8Mpc$, and the inclination angle $\theta_0=17^o$ \cite{Akiyama:2019cqa,Akiyama:2019brx,Akiyama:2019sww,Akiyama:2019bqs,Akiyama:2019fyp,Akiyama:2019eap}. The preliminary analysis of the EHT observations constrain the root-mean-square distance from the average radius of the black hole shadow as $\Delta C \lesssim 0.1$ which also named as circularity deviation, the axis ratio as $1 < D_{x} \lesssim 4/3$. Then, in Refs. \cite{Psaltis,EventHorizonTelescope:2021dqv}, the authors addressed that the shadow size of M87* should lie in the range as $R_a\approx3\sqrt{3}(1\pm0.17)M$. Many works have used these shadow observables $\Delta C$, $D_{x}$ and $R_a$ to constrain the parameters of corresponding black holes \cite{Afrin,Bambi:2019tjh,Afrin:2021wlj}. Inspired by these works, we consider the accelerating charged rotating black hole as the supermassive black hole in M87*, and use the EHT observations to give the constraints on the corresponding parameters.

To obtain the constraints, we should introduce the explicit definitions of the observables $\Delta C$, $D_{x}$ and $\psi_d$, respectively. The center of the shadow can be set as ($x_c=\frac{x_r+x_l}{2},y_c=0$). The boundary of the shadow can be expressed with the polar corrdinates ($\phi, R(\phi)$) as
\begin{eqnarray}
\phi=\tan^{-1}\Big(\frac{y}{x-x_c}\Big),~~
R(\phi)=\sqrt{(x-x_c)^2+y^2},
\end{eqnarray}
and the circularity deviation $\Delta C$ of the black hole shadow can be defined as \cite{Bambi:2019tjh}
\begin{eqnarray}
\Delta C=\frac{1}{\bar{R}}\sqrt{\int^{2\pi}_0(R(\phi)-\bar{R})^2d\phi},
\end{eqnarray}
with the average radius
\begin{eqnarray}
\bar{R}=\frac{1}{2\pi}\int^{2\pi}_0R(\phi)d\phi.
\end{eqnarray}
The axis ratio is expressed \cite{Kumar:2018ple,Banerjee:2019nnj}
\begin{eqnarray}
D_x=\frac{y_t-y_b}{x_r-x_l},
\end{eqnarray}
and the definition of the characteristic areal radius of the shadow curve can be expressed as \cite{Banerjee:2019nnj,Abdujabbarov:2015xqa}
\begin{eqnarray}
R_a=\sqrt{\frac{2}{\pi}\int^{r_{max}}_{r_{min}}\Big(y(r)\frac{dx(r)}{dr}\Big)}. 
\end{eqnarray}

\begin{figure}[htbp!]
\begin{center}
\subfigure[]{\label{AdetaC}
\includegraphics[width=0.232\textwidth]{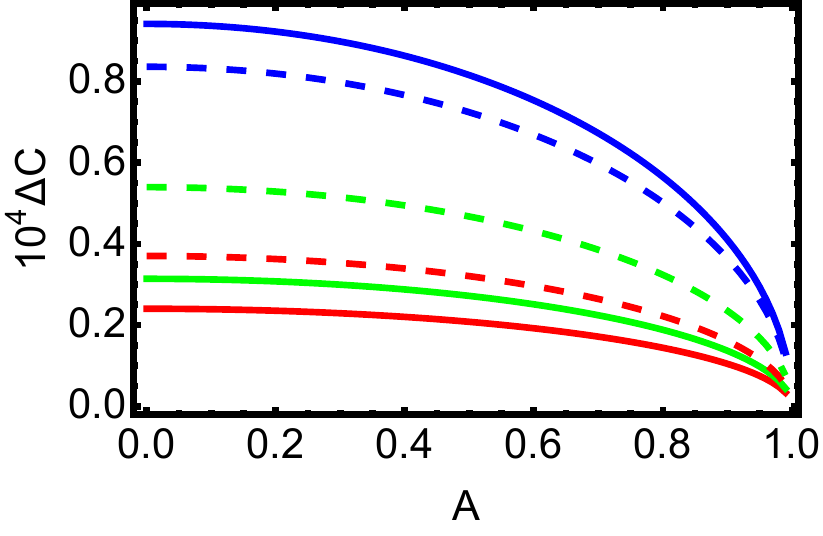}}
\subfigure[]{\label{ADx}
\includegraphics[width=0.232\textwidth]{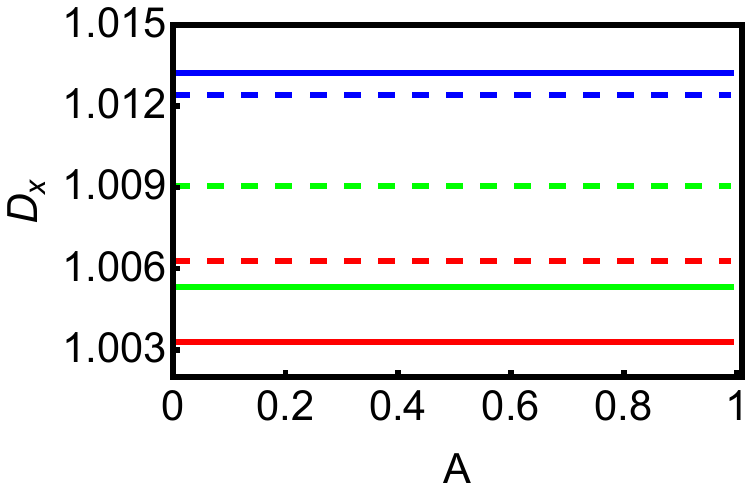}}
\end{center}
\caption{The density diagrams of the circularity $\Delta C$ and axial ratio $D_x$ with the ETH data $M=6.5\times10^9M_\odot$, $r_0=16.8Mpc$, and $\theta_0=17^o$. The solid lines are red ($a=0.6$), green ($a=0.7$), and blue ($a=0.86$)  with $e=0.5$. The dashed lines are red ($e=0.3$), green ($e=0.5$) and green ($e=0.76$) with $a=0.8$.}\label{f6}
\end{figure}

\begin{figure}[htbp!]
\begin{center}
\subfigure[]{\label{detaC}
\includegraphics[width=0.232\textwidth]{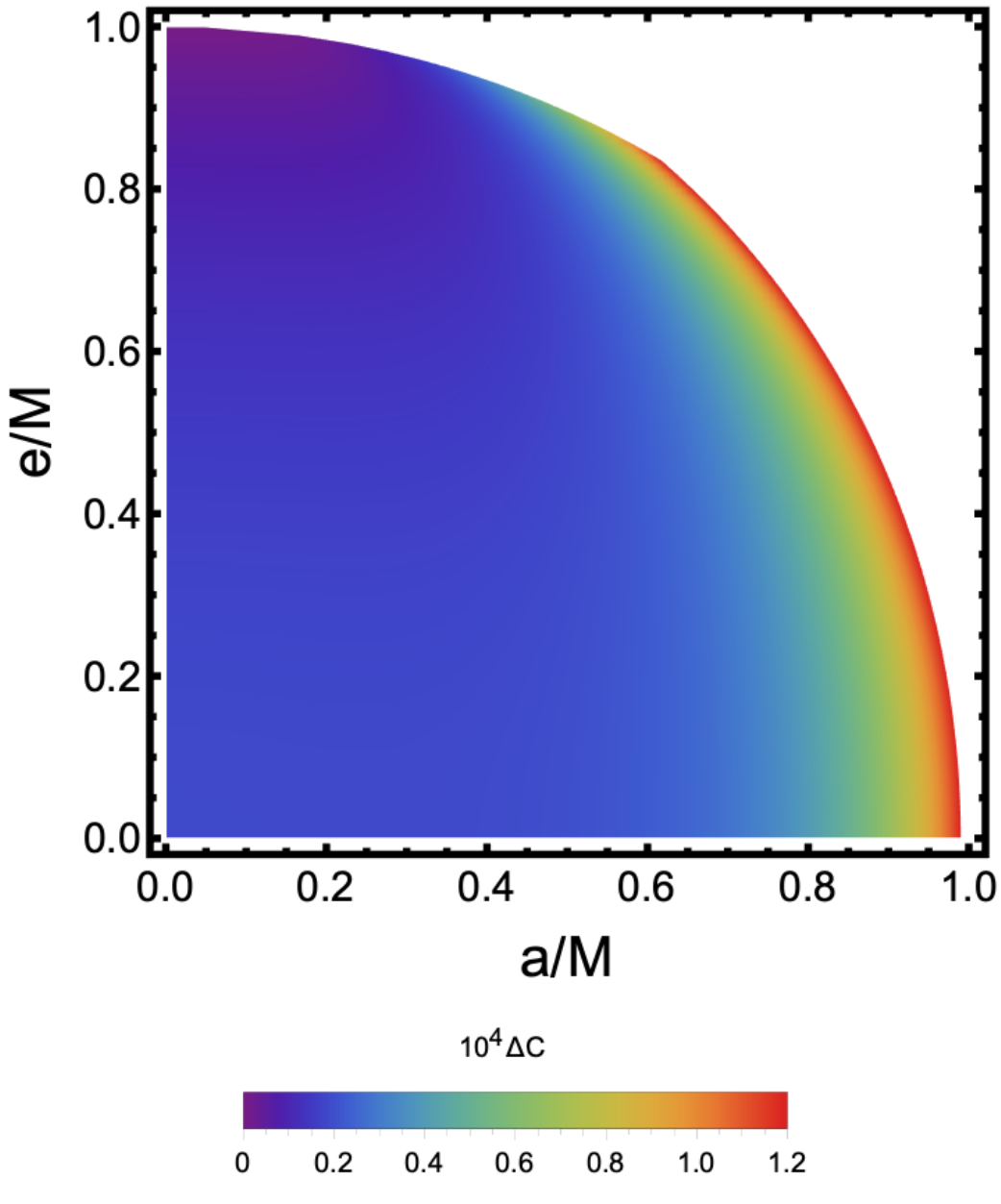}}
\subfigure[]{\label{aeDx}
\includegraphics[width=0.2325\textwidth]{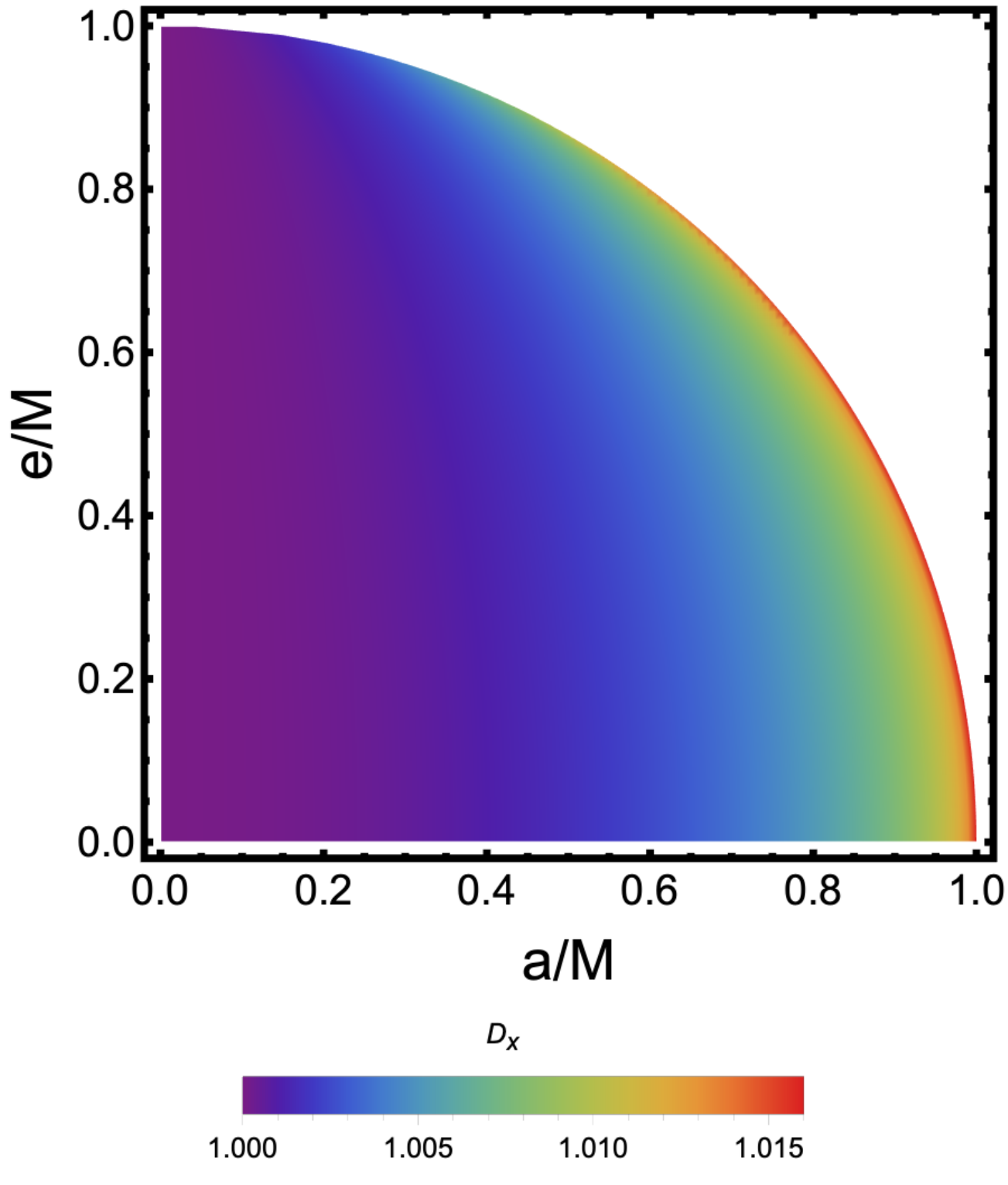}}
\end{center}
\caption{The density diagrams of the circularity $\Delta C$ and axial ratio $D_x$ in the $(a,e)$ plane with the ETH data $M=6.5\times10^9M_\odot$, $r_0=16.8Mpc$, and $\theta_0=17^o$. The parameter $A$ is set as  $A=0.01\frac{1}{r_0}$.} \label{f7}
\end{figure}

From the definitions of the obsverations, we can see that both the circularity deviation $\Delta C$ and the axis ratio $D_x$ are the relative obserations, which means the expansion or reduction of the outline for the shadow will not affect $\Delta C$ or $D_x$. {Roughly speaking, Fig. \ref{thffect} shows that with the accelerating factor $A$ increasing, the diagrams of the shadow for the accelerating KN black hole will scale down, which implies that the EHT observables $\Delta C, D_x$ might not impose limitations on the range of accelerating factor $A$. This conclusion gains further support from Fig. \ref{f6}, which confirms that the EHT observations $\Delta C \lesssim 0.1$ and $D_x \lesssim 4/3$ are insufficient to effectively constrain the accelerating factor $A$. Additionally, Fig. \ref{f6} reveals a diminishing trend in circularity deviation $\Delta C$ with increasing accelerating factor $A$, while the impact of accelerating factor $A$ on the axis ratio $D_x$ remains negligible. Furthermore, for a KN black hole with near-extreme behavior or large rotation parameter $a$, both $\Delta C$ and $D_x$ are more pronounced. Consequently, to establish constraints on the parameters of the accelerating KN black hole using the EHT observables $\Delta C$ and $D_x$, it is advisable to present the outcomes in the $(a,e)$ plane. From Fig. \ref{f7}, we can see that the whole parameters space of ($a,e$) satisfy the EHT observations $\Delta C\lesssim0.1$ and $D_x\lesssim4/3$, which indicates that it is impossible to make any constraint on the parameters of the accelerating KN black hole}. 

\begin{figure}[htbp!]
\begin{center}
\includegraphics[width=70mm]{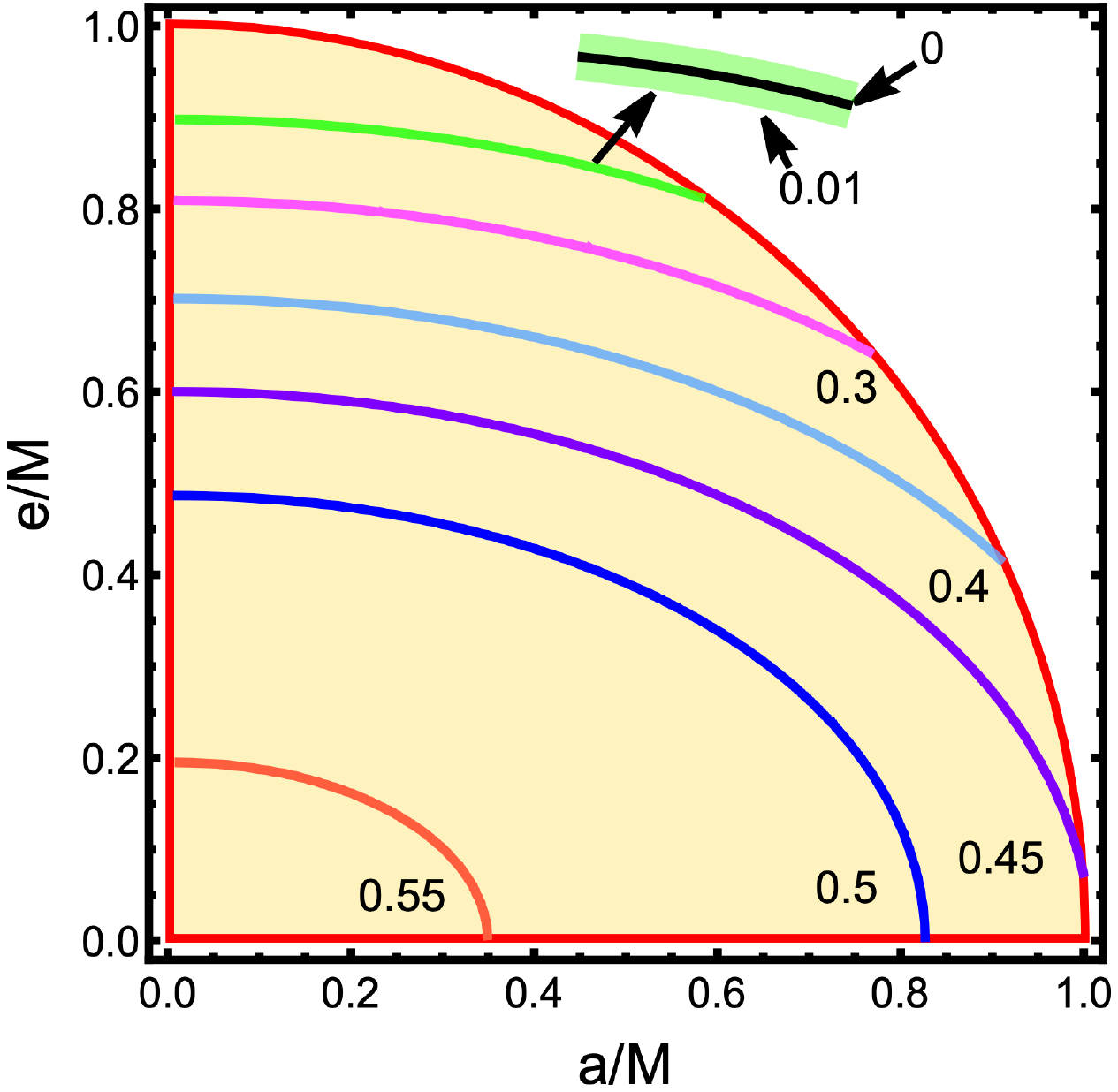}
\end{center}
\caption {The constraints of the ETH observation $4.31M \leq R_a\leq6.08M$ on the parameters ($a,e$). The real curves represent $R_a=4.31M$ with different accelerating factor $A$.} \label{xianzhiae}
\end{figure}

\begin{figure}[htbp!]
\begin{center}
\includegraphics[width=70mm]{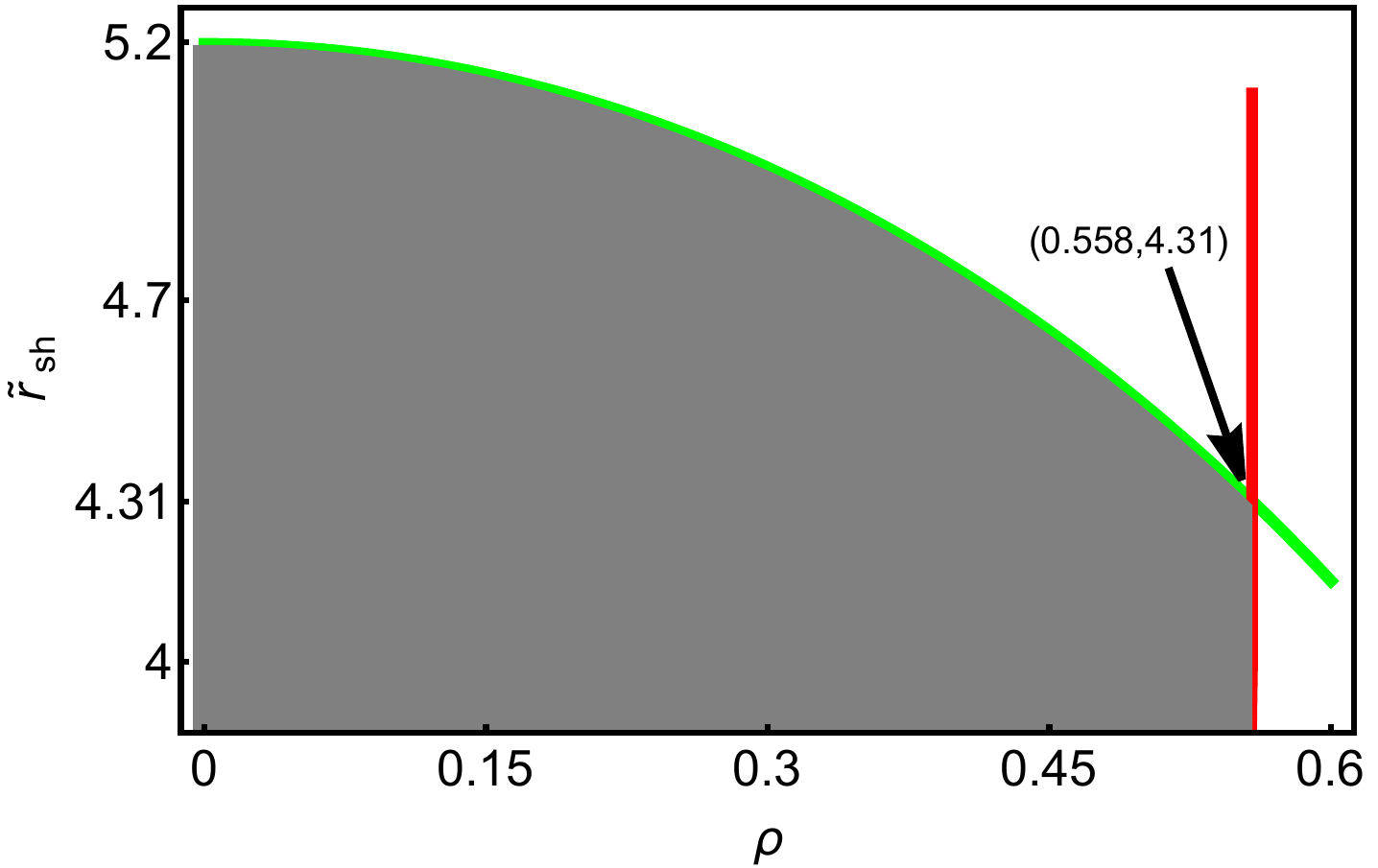}
\end{center}
\caption {The constraint of the ETH observation $4.31M \leq R_a\leq6.08M$ on the accelerating factor $A$ with the parameters $a/M=e/M=0$. {Here, we set $A=\rho \frac{r}{r_0}$.}}\label{xianzhiA}
\end{figure}

Figure \ref{xianzhiae} shows the constraints of the parameters for the black hole by the EHT observation with the shadow size $R_a$. All the real curves represent the lower bound $R_a=4.31M$, and the reasonable parameters with the constraints below these curves. The results illustrate that the bigger accelerating factor $A$, the stronger restriction on the rotating parameter $a$ and charge parameter $e$. From the extremely slow accelerating case, i.e., $Ar_0\leq0.01$, the charge parameter $e$ has maximum range as $e/M\in(0,0.9)$, which is consistent with the conclusion in \cite{EventHorizonTelescope:2021dqv}. While, according to the range of shadow size $R_{a}$, we can not make any constraints on the rotating parameter $a$.  Furthermore, we consider the maximum value range of the accelerating factor $A$ by setting the parameters as $a/M=e/M=0$, and the results shows that $Ar_0\in(0,0.558)$,  which is shown in Fig. \ref{xianzhiA}.


\section{Summary}\label{sec6}
In this paper, we concentrated on the shadow of the accelerating KN black hole, and analyzed the influence of the parameters on the shadow and its observables. The rotating parameter $a$, the charged parameter $e$, and the inclination angle $\theta_0$ affect the shadow with the qualitatively similar to that of Kerr-Newman black holes \cite{Perlick,Grenzebach,Tsukamoto,Xavier,Meng}. Besides, the size of the shadow for the accelerating KN black hole case will scale down with the accelerating factor $A$. 

Then, we analyzed the influence of the parameters on the shadow observables, i.e., the reference radius $R_s$ and the distortion $\delta s$. The rotating parameter $a$ dominates the change of the distortion $\delta s$, and the reference radius $R_s$ decreases with the charge parameter $e$, rapidly. The result shows that although the accelerating factor $A$ can affect the best viewing angle, which makes $R_s$ or $\delta s$ maximum, the deviations from $\theta_0=\frac{\pi}{2}$ are limited, i.e., the deviations are less than $1\%$. 

Finally, we assumed the accelerating KN black hole as the supermassive M87* black hole, and used the EHT observables to constrain the black hole parameters. We found that the whole parameters space ($a,e$) satisfy the EHT observables $\Delta C$ and $D_x$ constraints, which means that it can not rule out the supermassive M87* black hole is an accelerating KN black hole through the EHT observables $\Delta C$ and $D_x$. On the other hand, according to the observable shadow size $R_a$, the result shows that the bigger accelerating factor $A$ is, the smaller ranges of parameter $a$ and $e$ are. For an extremely slow accelerating case $(Ar_0\leq0.01)$, the ranges of rotating parameter $a$ and charge parameter $e$ are $a/M\in(0,1)$ and $e/M\in(0,0.9)$. The maximum range of the accelerating factor is $Ar_0\leq 0.558$ for a accelerating Schwarzschild case with $(a/M = e/M = 0)$.

\section*{Acknowledgements}
This work is supported in part by the National Natural Science Foundation of China (Grants No. 12147175, No. 12205129, No. 12175105, No. 11575083, No. 11565017), and the China Postdoctoral Science Foundation (Grant No. 2021M701529, No. 2021M700729).

%

\end{document}